\magnification=1200
\baselineskip=18truept

\line{\hfill RU-97-36}

\vskip 2truecm
\centerline{\bf Comments on ``Lattice Formulation of the Standard
Model'' by Creutz. et. al. }

\vskip 1truecm
\centerline{Herbert Neuberger}
\vskip .5truecm

\centerline {Department of Physics and Astronomy}
\centerline {Rutgers University, Piscataway, NJ 08855-0849}

\vskip 1.5truecm

\centerline{\bf Abstract}
\vskip 0.75truecm
The main construction of the
paper in the title is 
summarized using more standard particle physics
language. Also, a flaw is pointed out and
several objections to 
the more general thoughts expressed by Creutz et. al. are raised.

\vfill\eject

Creutz et. al. [1] consider a model in which every fermion of the
standard model is mirrored by an additional fermion differing only
in handedness. One extra right handed neutrino is added per
generation. The discussion is mainly restricted to a single generation.
Clearly, the set-up calls for $SO(10)$ GUT notation [2]. More specifically,
the chain $ SO(10) \rightarrow SU(4)_{PS}  (\approx SO(6)) 
~ \otimes ~ [SU(2)_L ~\otimes ~ SU(2)_R ] (\approx SO(4)) $ is relevant,
with $PS$ referring to Pati-Salam. The one generation $16$ decomposes into
$(4,2,1) \oplus (\bar 4 , 1, 2) \equiv \Psi^L \oplus \Psi^R$. 
The central object in [1] is a dimension six Baryon number violating 
operator. For a single generation there are four possibilities
[3]. Creutz et. al. make what amounts to a choice of ${\cal O}^{(3)}$
in Weinberg's notation [3]. Using the notation of [1] for the fermions
(except for the explicit handedness), introduce the scalar fields
$\Phi^L_{\alpha \beta} \equiv 
\Psi^L_{\alpha i s} \Psi^L_{\beta j t} 
\epsilon_{ij} \epsilon_{st}= -\Phi^L_{\beta \alpha}$ 
and similarly for $L$ replaced by $R$. $\Phi^{L,R}$ are singlets
under both the left and right $SU(2)$'s. For $(\alpha\beta)$ 
restricted to $SU(3)$-color $\Phi^L_{\alpha\beta}$ 
is a $\bar 3$ and for  $\beta=4$ it is a $3$. 
Under $SU(4)_{PS} (\approx SO(6))$ $\Phi^L$ is an antisymmetric rank
two tensor (six component vector). $\Phi^L$ can be decomposed
into selfdual and anti-selfdual tensors $\hat\Phi^{L\pm}_{\alpha\beta}=
\Phi^L_{\alpha\beta} \pm {1\over 2} \epsilon_{\alpha\beta\gamma\delta}
\Phi^L_{\gamma\delta}$.
The Baryon number violating operator 
is a scalar under $SU(4)_{PS}\otimes SU(2)_L \otimes SU(2)_R$ given
by
$
V_L =\epsilon_{\alpha\beta\gamma\delta} \Phi^L_{\alpha\beta}
\Phi^L_{\gamma\delta}$. 
In terms of $\hat\Phi^{L\pm}$, $V_L$ is diagonal.
Similar definitions give $V_R$. 
The main idea is to add to the Lagrangian a
term $ g[V_L +V_R +h.c.]$ for each generation and only
for the mirrors. $V_L$ could be replaced by coupling
an auxiliary field $\phi^L_{\alpha\beta}$ to $\Phi^L_{\alpha\beta}$,
making the Lagrangian bilinear in the $\Psi$'s.

First, in the spirit of the
``Yukawa approach'' to the regularization of chiral
gauge theories ([1]), the theory is studied in the
absence of gauge fields. On the lattice, if $g$ is strong enough,
and one ignores coupling between the mirrors and 
the ordinary fermions, only the mirrors become massive and their
masses are large. It is hoped that this will persist to 
weaker $g$'s and that the theory stay in a symmetric 
phase $(<\Phi^{L,R}_{\alpha\beta} > =0)$. Then, one hopes that when
the ordinary $SU(3)\otimes SU(2)_L \otimes U(1)_Y$ gauge
interactions are turned on, the ordinary Higgs mechanism
would be {\it necessary} to render massive the $W$'s, the $Z$ 
and the fermions.

The mirrors and the ordinary fermions interact  
via $L-R$ couplings through a chain 
of heavy Dirac fermions, 
along a fictitious fifth dimension as suggested by Kaplan [4]. On the
lattice this is a finite segment and it is hoped that it will
prevent the mirror masses from feeding down to the ordinary fermions.
Some of the anomalous 
global symmetries carried by ordinary particles
in the continuum are not exact on the lattice since the mirrors do not
have strictly infinite masses and the presence
of $V_L + V_R$ is felt. This method of eliminating unwanted
symmetries is due to Eichten and Preskill [5].  It is hoped that in the
presence of an $SU(2)_L$ instanton the correct 't Hooft vertex [6]
will appear in the continuum limit independently of the strength
of $g$.

Actually, the model of [1] is invariant under a 
discrete $Z_3 \otimes Z_3$ 
symmetry: $u_R , d_R \rightarrow z_R u_R , z_R d_R$
and $u_L , d_L \rightarrow z_L u_L , z_L d_L$ ($z_{R,L}^3 =1$),  
independently for each generation. The true standard model
would not obey this symmetry due to color-instantons. Full
compliance with the Eichten-Preskill guidelines would forbid
this discrete global symmetry. 

Although the construction of [1] is specific to 
the standard model the authors make some general remarks.
As a chiral gauge theory the standard model is relatively simple,
as the single ``dangerous'' group is a $U(1)$. Moreover,
the related coupling isn't asymptotically free. It is
unclear then how much can be inferred about the general problem
of regulating an asymptotically free 
non-abelian chiral gauge theory in four dimensions.  
The authors view as the main weakness of their approach 
the questionable 
existence of the appropriate phase in the ungauged model.
``Such a situation would cast serious doubts on any construction
of chiral gauge theories'' they say referring to the possibility of
the desired phase being ``squeezed out''. In view of the many
arbitrary choices they made (one particular Baryon number violating
operator, a left-right symmetric model prior to gauging, a
mirror-Yukawa approach [7], and the generic failure rate in 
simpler Yukawa models) it is difficult to justify the ``any''
in their statement. 

The most confusing statements in [1] are made comparing
their proposal to the ``overlap'' [8]. The latter is a general 
scheme which can be interpreted as treating an infinite number
of lattice fermions. However, there is nothing infinite in the
overlap regularization. Nevertheless, 
Creutz et. al. conclude that their approach
is ``cleaner in that gauge invariance is exact, all infinities
are eliminated, and the requirement of anomaly cancelation
is manifest''. 

The manifest requirement for anomaly cancelation
is justified elsewhere in [1]: it simply means that
the charged Lepton charge has to be 3 times the quark charge to
ensure gauge invariance of the vertex. 
Anomaly cancelation is not {\it directly} required, but
just happens to hold because, for example, we don't have
more charged fermions that don't participate in the vertex. 
Working in an $SO(10)$ scheme guaranteed anomaly cancelation 
from the start. 
As a matter of fact, 
a 't Hooft vertex is explicitly used in the overlap 
in order to 
completely define the model (section 5.2 of [8]). 
Thus, the overlap is similar to [1] 
in this respect. But, [1] has built 
in $\Delta B =\Delta L =1$ for Baryon and Lepton number violations.
Instantons only allow $\Delta B =\Delta L =3$ and the overlap
has clear preference for these processes. Also, the overlap
would produce no unwanted global symmetries, global or discrete.

If one applied the overlap to the standard model one would
have to integrate over a degree of freedom $e^{i\theta}$
for each site on the four dimensional lattice. 
The $\theta$'s can be interpreted as a gauge
degree of freedom - somewhat similar to a longitudinal photon. 
It is hoped [8] that the integration over $\theta$
would restore $U(1)$ invariance without
adding extra nonlocal terms, similarly to [9]. 
If the $U(1)$ were anomalous
this cannot happen [10]. So, anomaly cancelation plays
a much more intrinsic role in the overlap than in [1]. 
In [1], similarly to the Yukawa approach,
the phase structure of the ungauged theory is deemed crucial.
This phase structure cannot be influenced by which group
we intend to gauge, so the phases are hardly dependent
on anomaly cancelation. It is {\it a priori} possible to
find the ``right'' phase, 
while some particular gauging still shouldn't work because
anomalies do not cancel. In practice, to implement
the proposal of [1], an integration
over some auxiliary fields, like $\phi^{L,R}_{\alpha\beta}$, 
will always be needed. In the overlap, the $\theta$'s
at least do not carry color or charge. 

The overlap definition includes the above 
gauge averaging, so there is an almost tautological gauge invariance.
Thus, exact gauge invariance isn't really an issue. 
What does matter though are the ``hopes'' about 
the degrees of freedom surviving the continuum 
limit: In the overlap one hopes
that with anomaly cancelation the variables $\theta$ will stay
massive and decouple as in [9]. Creutz et. al. hope that the
whole slew of mirrors, (possibly bound into $\Phi$'s ?) will 
decouple. If, for example, $\Phi$ acquires an expectation value
Creutz et. al.'s model would 
loose ordinary confinement. It should sound strange that
the strong interactions are put in danger as a result of trying
to gauge the weak $U(1)$. 
In the overlap, $SU(3)$-color would be well isolated. 

The overlap has been subjected to a real, albeit modest, test [11]:
In an exactly soluble abelian 
two dimensional chiral model the 't Hooft vertex was shown
to come out correctly. Thus, the hopes about the overlap
have been shown to come true at least in one non-trivial 
instance. Nothing has been reported that compares even remotely
for any example of the Yukawa approach, [1] included. Creutz et. al.
describe the model in [11] as employing a ``tricky twist''
because the 't Hooft vertex contains a derivative. 
In any dimension, the vertex contains 
one fermion field for each fermionic
zero mode, all at the same point in Euclidean space.
Quite often, derivatives will be needed to make
the vertex a Lorentz scalar. 
In four dimensions this happens, for example, in N=1 supersymmetric 
pure SU(2) gauge theory [12].
A proposal that cannot work when there are derivatives in the
't Hooft vertex is unreasonably restricted. 
While two dimensions differ substantially from four, 
numerical, dynamical, fully
non-perturbative tests
of any proposal are practical only in two dimensions (excluding
odd dimensions) at present. The issues related to chirality specifically 
are reasonably similar in all even dimensions and it is difficult
to accept that a proposal could work in four dimensions but fail
in two. 

In the closing paragraph of [1] it is suggested that a success
of their proposal would justify several other approaches,
among them the overlap. Indeed, the fifth dimension,
being of arbitrary length, 
could be taken to infinity yielding 
something similar to the overlap, now with
trivial gauge averaging. But, the
overlap, when interpreted as a theory containing an infinite
number of fermions also contains an infinite subtraction
of the effective action due to the infinitely many heavy
particles. Including such a subtraction will always
mar a ``pure'' action interpretation where ghost 
fields are excluded. 
A success of the proposal of [1] cannot justify some of the other
proposals they mention: unlike the overlap, most of them 
do contain explicit infinities and ignore instantons.

In conclusion, the approach proposed in [1] is too special 
to allow drawing any general lessons about 
defining non-perturbatively chiral gauge theories, or
about other approaches to the problem. In
its present form the construction of [1]
has a potential flaw, an unwanted
global discrete symmetry. In practice, 
it is unlikely we'll know in the foreseeable future whether the 
full four dimensional proposal
of [1] works or not. Even if the desired phase were
to be ruled out in some simpler variant, we couldn't attach 
a general meaning to the finding. Since the authors
discount two dimensional tests, the prospects for
any objective evidence of success in their approach
are slim.

\bigskip
\centerline{\bf  Acknowledgments.}
\medskip

This research was supported in part by the DOE under grant \#
DE-FG05-96ER40559.

\bigskip
\bigskip
\centerline{\bf  References.}
\medskip

\item{[1]} M. Creutz et. al., hep-lat/9612017.
\item{[2]} H. Georgi, D.V. Nanopoulos, Nucl. Phys. B155, 52 (1979).
\item{[3]} S. Weinberg, Phys. Rev. Lett. 43,  1566 (1979);
F. Wilczek, A. Zee, Phys. Rev. Lett. 43,  1571 (1979);
L. F. Abbott, M. B. Wise, Phys. Rev. D22,  2208 (1980).
\item{[4]} D. Kaplan, Phys. Lett. B288, 342 (1992).
\item{[5]} E. Eichten, J. Preskill, Nucl. Phys. B268, 179 (1986).
\item{[6]} G. 't Hooft, Phys. Rev. Lett. 37, 8 (1976).
\item{[7]} I. Montvay, Phys. Lett. 199B, 89 (1987).
\item{[8]} R. Narayanan, H. Neuberger, Nucl. Phys. B443, 305 (1995).
\item{[9]} D. Foerster, H. B. Nielsen, M. Ninomiya, 
Phys. Lett. 94B, 135 (1980).
\item{[10]} J. Preskill, Ann. Phys. 210, 323  (1991). 
\item{[11]} R. Narayanan, H. Neuberger, hep-lat/9609031, Phys. Lett. B,
to appear;\hfill\break  Y.  Kikukawa, R. Narayanan, 
H. Neuberger, hep-lat/9701007, Phys. Lett. B, to appear; hep-lat/9705006.
\item{[12]} A. I. Vainshtein, V. I. Zakharov, JETP Lett. 35, 323 (1982).

\vfill\eject\end